%% file: Fig-of-merit-PRApp_v14d-clean.tex
\documentclass[english,prl,superscriptaddress,reprint]{revtex4-1}
\usepackage[T1]{fontenc}
\usepackage{verbatim}
\usepackage{amsmath}
\usepackage{amssymb}
\usepackage{graphicx}
\usepackage{esint}

\makeatletter
\usepackage{amsfonts}
\usepackage{graphicx}
\usepackage{xcolor}
\definecolor{darkblue}{rgb}{0,0,0.5} 
\usepackage{transparent}

\makeatother

\usepackage{babel}
\begin{document}
\selectlanguage{english}

\author{Emil Zeuthen}

\email{zeuthen@nbi.ku.dk}

\affiliation{Niels Bohr Institute, University of Copenhagen, DK-2100 Copenhagen,
Denmark}
\affiliation{Institute for Theoretical Physics \& Institute for Gravitational
Physics (Albert Einstein Institute), Leibniz Universit\"{a}t Hannover,
Callinstra{\ss}e 38, 30167 Hannover, Germany}

\author{Albert Schliesser}
\author{Anders S. S{\o}rensen}

\affiliation{Niels Bohr Institute, University of Copenhagen, DK-2100 Copenhagen,
Denmark}

\author{Jacob M. Taylor}

\affiliation{Joint Quantum Institute, University of Maryland/National Institute of Standards and Technology, College Park, Maryland 20742, USA}
\affiliation{Joint Center for Quantum Information and Computer Science, University of Maryland, College Park, Maryland 20742, USA}

\title{Figures of merit for quantum transducers}
\begin{abstract}
Recent technical advances have sparked renewed interest in
physical systems that couple simultaneously to different parts of
the electromagnetic spectrum, thus enabling transduction of signals between vastly different frequencies at the level of single photons. Such hybrid
systems have demonstrated frequency conversion of
classical signals and have the potential of enabling quantum state
transfer, e.g., between superconducting circuits  and traveling optical signals.
This article describes a simple approach for the theoretical characterization of the performance of quantum transducers.
Given that, in practice, one cannot attain
ideal one-to-one quantum conversion, we explore how well the transducer performs in scenarios ranging from classical signal detection to applications for quantum information processing. While the performance of the transducer depends on the particular application in which it enters, we show that the performance can be characterized by defining two simple parameters: the signal transfer efficiency~$\eta$ and the added noise~$N$.
\end{abstract}
\maketitle

\section{Introduction}
Interconversion of signals between electrical and optical domains is a crucial task for modern information processing and communication.
The impressive technological advances in our ability to control individual quanta have spurred the development of similar devices operating at the quantum level.
In the quantum setting, the frequency conversion of individual photons such as those in superconducting circuits in a dilution refrigerator to outgoing, optical photons may enable an optically connected ``quantum internet'' based on superconducting quantum computers~\cite{Kimble2008,Stannigel2010}. A related application is quantum-limited detection of electrical signals~\cite{Taylor2011,Zhang2015} as in, e.g., nuclear magnetic resonance readout~\cite{Takeda2017} or radio astronomy.
Such quantum applications put much stricter requirements on the conversion process, which is being actively pursued in a number of different physical setups~\cite{Takeda2017,Bochmann2013,Hill2012,Pitanti2015,Vainsencher2016,Balram2016,Andrews2014,Unterreithmeier2009,Schmid2014,Schmid2014,Andrews2014,Bagci2014,Lecocq2016,Fink2016,Takeda2017,Bochmann2013,Vainsencher2016,Balram2016,Hill2012,Sliwa2015,Hisatomi2016,Javerzac-Galy2016}.
In the quantum setting the action of the transducers is now constrained by the microscopic laws of quantum mechanics, most notably unitarity. This entails that the performance of quantum transducers cannot be fully described  by the same figures of merit as classical transducers and hence a new performance metric has to be developed for quantum transducers.

While transduction, broadly defined, is ubiquitous in applications of physics and engineering, we will here restrict our focus to transducers that perform frequency conversion of one itinerant mode to another in the quantum regime. 
A promising candidate mechanism for such transduction is provided by optomechanics~\cite{Aspelmeyer2013,CavityOptomechSpringer}, which relies on radiation pressure to efficiently couple an optical field to a high-$Q$ mechanical mode.
This can be implemented in various systems, e.g., a mechanically compliant photonic crystal cavity~\cite{Bochmann2013,Hill2012,Pitanti2015,Vainsencher2016,Balram2016} or a membrane embedded in a standard Fabry-P\'{e}rot cavity~\cite{Andrews2014}.
Since, in principle, electromagnetic radiation from any part of the spectrum can exert a force on mechanical objects, electromechanical coupling to, e.g., radio-frequency and microwave fields can likewise be engineered. Mechanisms for implementing this include the
Kelvin polarization force on a dielectric~\cite{Unterreithmeier2009,Schmid2014}, the quasi-electrostatic force on a conductor~\cite{Schmid2014,Andrews2014,Bagci2014,Lecocq2016,Fink2016,Takeda2017}, and piezoelectricity~\cite{Bochmann2013,Vainsencher2016,Balram2016}.
Combining the above techniques to attain simultaneous opto- and electromechanical couplings for a single mechanical mode permits it to act as an efficient intermediary between radiation modes of vastly different frequencies, as investigated in a variety of theoretical proposals~\cite{Stannigel2010,Regal2011,Safavi-Naeini2011,Taylor2011,Zhang2015,Wang2012,Tian2012,Barzanjeh2012,Tian2015,Ondrej2017}. 
Transduction based on these techniques has been realized with input and output modes respectively in the radio-frequency~\cite{Bagci2014,Takeda2017} or microwave domain~\cite{Bochmann2013,Andrews2014,Vainsencher2016} and the optical domain, and between two frequencies both within either the optical~\cite{Hill2012} or microwave~\cite{Lecocq2016} domain.
In parallel pursuits, quantum-level conversion between microwave frequencies has been achieved in superconducting circuits~\cite{Sliwa2015}, and it has been proposed to mediate electro-optic quantum transduction via an erbium-doped crystal~\cite{OBrien2014,Williamson2014}, a ferromagnetic magnon~\cite{Hisatomi2016}, a spin ensemble~\cite{Blum2015}, or by direct electro-optic coupling~\cite{Tsang2011,Javerzac-Galy2016}.

Regardless of the physical system underlying it, the action of a realistic transducer amounts to a signal being (partially) transferred from one mode to another along with some added noise. How imperfect transduction  influences the performance of various quantum application remains to be elucidated and calls for defining  transduction metrics characterizing the performance. 
Establishing transduction metrics identifying feasible applications for a given transducer could serve as a road map for its optimization towards a particular application as well as a means to compare different transducers based on different physical systems.

In this article, we consider the usual metrics for amplifiers~\cite{Clerk2010} -- efficiency $\eta$ and added noise $N$ -- and use them to describe the performance of linear transducers for quantum information tasks including single photon measurements and high fidelity transfer of Fock state superpositions. We show that the non-trivial problem of optimizing a quantum transducer is intuitively captured by considering the fate of a single photon entering the transducer, yielding one of the following outcomes: the input photon comes out (quantum transduction), no photon comes out (loss), or other photons come out (noise and/or amplification). Striking the right balance between these competing processes given practical constraints is the central challenge of quantum transduction. Here we show that the optimal trade-off is application-dependent, but regardless of the application the performance is always characterized by the two parameters $\eta$ and $N$. This is in contrast to the classical transduction literature, in which only the added noise relative to the input 
is typically relevant.

\section{Scattering matrix formulation of linear transduction}
\begin{figure}
\centering
\def\svgwidth{\columnwidth}
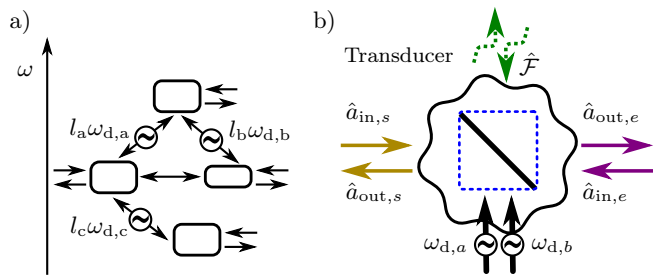
\caption{Generic transduction scenario. a) Example of transducer topology in frequency space. The various modes (boxes) of the transducer are coupled as indicated by the double-headed arrows. Coupling of non-degenerate modes requires an AC drive field (frequency $\omega_{\text{d,}i}$) to bridge the frequency gap by an integer number of drive frequencies $l_{i} \omega_{\text{d,}i}$. In general, each mode is subject to signal and noise inputs (incoming single-headed arrow) as well as readout and loss (outgoing single-headed arrow). b) A transducer is driven by harmonically varying drive frequencies $\omega_{\text{d},i}$ which connect different frequency components. In an idealized limit the transducer acts as a beam splitter  transforming an input signal $\hat{a}_{\text{in,s}}$ at a certain frequency to an output signal $\hat{a}_{\text{out,e}}$ at a different frequency by addition and/or subtraction of the drive frequencies. A finite transduction efficiency $\eta<1$ leads to admixture of noise $\hat{\mathcal{F}}$ from the other port of the beam splitter.\label{fig:Beam-splitter_RSB-transducer}}
\end{figure}

Linear transducers use oscillating bias fields to parametrically modulate an incoming field and thereby connect different frequency components as illustrated in Fig.~\ref{fig:Beam-splitter_RSB-transducer}a. 
Here we are interested in the linearized regime where a weak (bosonic) field is transduced to a different frequency through modulation of a nonlinear element with a much stronger classical bias field. In this case, the Hamiltonian can be truncated at the second order in the involved field operators
for weak signals. As a result, the solution to the Heisenberg-Langevin equations in the Fourier domain will take the form 
\begin{equation}
\vec{A}_{\text{out}}(\Omega)=\int d\Omega' \mathbf{S}(\Omega,\Omega')\vec{A}_{\text{in}}(\Omega'),\label{eq:scat-matrix-gen}
\end{equation}
where $\vec{A}_{\text{in(out)}}(\Omega)$ is a vector containing the input (output) frequency ($\Omega$) components of the annihilation and creation operators $\hat{a}_i$ and $\hat{a}_i^\dagger$ of all the involved modes, including decay channels.
The action of the transducer is captured by the scattering matrix $\mathbf{S}(\Omega,\Omega')$, which describes how it maps the input frequency components to the output; its elements are constrained by unitarity so as to preserve the commutation relations of the itinerant fields~\cite{Caves1982}.
We emphasize that Eq.~\eqref{eq:scat-matrix-gen} represents the generic description of a quantum transducer operating in the linear regime. Regardless of the physical system being investigated, the description of it will  be of the form~\eqref{eq:scat-matrix-gen}. Once the scattering matrix $\mathbf{S}$ has been determined, a remaining question is how well the transducer performs for various applications. This is the question addressed in this article.

For specificity, we assume the drive amplitude of the bias fields to be independent of time~\cite{Safavi-Naeini2011},
though our approach can be extended to time-varying amplitudes~\cite{Tian2010,Wang2012,Tian2012,McGee2013} or detunings~\cite{Zhang2015} or using quantum teleportation~\cite{Barzanjeh2012}.
This assumption results in a well-known form of the matrix $\mathbf{S}(\Omega,\Omega')$ (see Appendix~\ref{sec:app-scat}). We further specialize to the scenario where all field operators can be divided into a finite number of (narrow) frequency bands of center frequencies $\omega_{0,m}$ which are separated by an integer number $l$ of one of the drive frequencies, $\omega_{0,m}-\omega_{0,m'}=l\omega_{\text{d,}i}$. In this case it is natural to express the scattering relation (\ref{eq:scat-matrix-gen}) between $\vec{A}_{\text{in(out)}}(\Omega)$ in terms of slowly-varying vector components $\hat{a}_{\text{in(out),}m}(\Omega)$ where the frequency $\Omega$ is defined relative to the band center $\omega_{0,m}$.
Given these assumptions, the scattering matrix takes a simple form dictated by energy conservation in the interaction between the slowly-varying operators: Excitation of well-defined frequency in the output field $\hat{a}_{\text{out,}m'}(\Omega)$ can only arise from either the annihilation of an excitation of the same (slowly-varying) frequency from one of the input ports $\hat{a}_{\text{in,}m}(\Omega)$ or the creation of an excitation of frequency $-\Omega$ corresponding to $\hat{a}^{\dagger}_{\text{in,}m}(-\Omega)$. These processes are accompanied by the absorption and/or creation of drive field photons to ensure energy conservation in the lab frame.
By analogy with the field of optomechanics we will refer to slowly-varying frequency components $\Omega >0$ ($\Omega <0$) of a given port as its upper (lower) sideband.

Focusing on the element of $\vec{A}_{\text{out}}$ corresponding to the ``exit'' port, $\hat{a}_{\text{out,e}}$, and considering a single-tone input at the upper sideband of the ``signal'' port, $\hat{a}_{\text{in,s}}(\Omega>0)$, we have the scattering relation
\begin{equation}
\hat{a}_{\text{out,e}}(\Omega)=\begin{cases}
U_{\text{s}}(\Omega)\hat{a}_{\text{in,s}}(|\Omega|)+\hat{\mathcal{F}}(\Omega) & \text{for }\Omega>0\\
V_{\text{s}}(\Omega)\hat{a}_{\text{in,s}}^{\dagger}(|\Omega|)+\hat{\mathcal{F}}(\Omega) & \text{for }\Omega<0
\end{cases},\label{eq:scat-rel}
\end{equation}
where
\begin{equation}
\hat{\mathcal{F}}(\Omega)\equiv \sum'_{m} \left[U_{m}(\Omega)\hat{a}_{\text{in,}m}(\Omega)+V_{m}(\Omega)\hat{a}_{\text{in,}m}^{\dagger}(-\Omega)\right]
\label{eq:F-noise_def2}
\end{equation}
is a stationary noise operator accounting for all other contributions to $\hat{a}_{{\rm out,e}}(\Omega)$; the prime on the sum~(\ref{eq:F-noise_def2}) excludes the upper-sideband contribution of the identified input mode `s' which was split out in Eq.~(\ref{eq:scat-rel}).
 The transducer will in general mix annihilation and creation operators of the input fields as implied by Eqs.~(\ref{eq:scat-rel},\ref{eq:F-noise_def2}).
The negative frequency components arising here should be understood in the rotating frame $\omega_{0,m}$ (for frequency band $m$) and are meaningful insofar as $\Omega>-\omega_{0,m}$.

\section{Example transducer}\label{sec:example}
\begin{figure}[t]
\begin{center}
\def\svgwidth{\columnwidth}
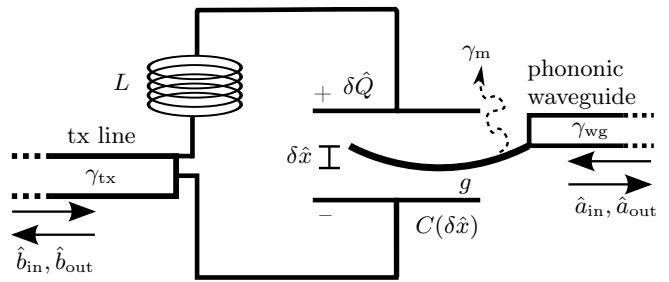
\caption{Electro-mechanical transducer linking the traveling fields $\hat{b}_{\text{in/out}}$ of an electrical transmission line (coupling rate $\gamma_{\text{tx}}$) with the traveling fields $\hat{a}_{\text{in/out}}$ of a phononic waveguide (coupling rate $\gamma_{\text{wg}}$). The link is provided by the internal modes of the transducer, an $LC$ electrical resonance with charge fluctuations $\delta \hat{Q}$, and a vibrational mode of a mechanical element with position fluctuations $\delta \hat{x}$ modulating the circuit capacitance $C(\delta \hat{x})$. The resulting coupling rate $g$ between the internal modes is enhanced by an AC drive (not shown) bridging their respective frequency scales. We include an additional, incoherent mechanical damping rate $\gamma_{\text{m}}$ to account for, e.g., friction.}
\label{fig:example}
\end{center}
\end{figure}

As an example of the formalism presented in the previous section, let us consider an electro-mechanical system comprising an $LC$ electrical circuit whose capacitance $C$ depends upon the position $x$ of a mechanical resonator mode~\cite{OConnell2010,Massel2012,Pirkkalainen2013,Palomaki2013,Reed2017}. 
This system can serve as a transducer between traveling electrical and mechanical fields, e.g., by coupling the circuit to a transmission line and the mechanical mode to a phononic waveguide
as will be considered here (see Fig.~\ref{fig:example}).
The transduction can be engineered by applying a harmonic drive tone of frequency $\omega_{\text{d}}$ to the circuit, so that an oscillating charge $\bar{Q}(t)=\bar{Q}_{0}e^{i\omega_{\text{d}}t}+\text{c.c.}$ is induced on the capacitor, providing an enhanced coupling between the fluctuations $\delta \hat{Q},\delta \hat{x}$ around the ensuing steady-state configuration of the capacitor charge $\hat{Q}=\bar{Q}+\delta \hat{Q}$ and mechanical position $\hat{x}=\bar{x}+\delta \hat{x}$~\cite{beta}.
We assume for specificity that the (steady-state) $LC$ resonance frequency $\bar{\omega}_{\text{LC}}=1/\sqrt{LC(\bar{x})}$ is in the GHz domain ($L$ is the inductance of the circuit) whereas the mechanical resonance $\bar{\omega}_{\text{m}}$ is in the MHz range; in this case, effective electro-mechanical transduction can be achieved by choosing a drive frequency $\omega_{\text{d}}\sim\bar{\omega}_{\text{LC}} - \bar{\omega}_{\text{m}}$ to bridge the two frequency scales. It is useful to introduce annihilation and creation operators for the electrical and mechanical degrees of freedom $\delta \hat{Q}=Q_{\text{zpf}}(\hat{b}+\hat{b}^{\dagger})/\sqrt{2}$ and $\delta \hat{x}=x_{\text{zpf}}(\hat{a}+\hat{a}^{\dagger})/\sqrt{2}$ with zero-point amplitudes $Q_{\text{zpf}}\equiv \sqrt{\hbar/(L \bar{\omega}_{\text{LC}})}$, and $x_{\text{zpf}}\equiv \sqrt{\hbar/(m \bar{\omega}_{\text{m}})}$; their canonical conjugate variables are the magnetic flux $\delta \hat{\phi}=-i(\hbar/Q_{\text{zpf}})(\hat{b}-\hat{b}^{\dagger})/\sqrt{2}$ and the mechanical momentum $\delta \hat{p}=-i(\hbar/x_{\text{zpf}})(\hat{a}-\hat{a}^{\dagger})/\sqrt{2}$, whereby we have the standard commutation relations $[\hat{a},\hat{a}^\dagger]=1=[\hat{b},\hat{b}^\dagger]$.
This permits us to write the linearized elecro-mechanical interaction Hamiltonian as
\begin{eqnarray}
\hat{H}_{\text{EM}} &=& \hbar (g/2) (\hat{b}+\hat{b}^{\dagger}) (\hat{a}+\hat{a}^{\dagger}) (e^{-i\omega_{\text{d}}t}+e^{i\omega_{\text{d}}t})\nonumber\\
&\approx& \hbar (g/2) (\hat{b}e^{i\omega_{\text{d}}t}+\hat{b}^{\dagger}e^{-i\omega_{\text{d}}t}) (\hat{a}+\hat{a}^{\dagger}),\label{eq:H-EM}
\end{eqnarray}
for a suitable coupling rate $g$ 
which scales linearly with the drive amplitude $\bar{Q}_{0}$ ($g$ is assumed real without loss of generality).
In the second line of Eq.~\eqref{eq:H-EM}, we have exploited the frequency scale separation $\omega_{\text{d}}\gg \bar{\omega}_{\text{m}}$ to neglect the rapidly varying terms $\propto\hat{b}e^{-i\omega_{\text{d}}t}(\hat{a}+\hat{a}^{\dagger})+\text{H.c.}$ The resulting coupling involves only the rotating frame circuit operator $\hat{b}e^{i\omega_{\text{d}}t}+\text{H.c.}$, whereas the mechanical operator $\hat{a}+\text{H.c.}$ enters in its ``lab'' frame representation; this asymmetry in reference frame entails (in general) a 2-to-1 folding of the electrical input spectrum around the drive frequency $\omega_{\text{d}}$ as it is transduced to the mechanical frequency domain, thus exemplifying the structure of the generic Eqs.~(\ref{eq:scat-rel},\ref{eq:F-noise_def2}).

The full Heisenberg-Langevin equations of motion for the electro-mechanical system follow from combining the interaction Hamiltonian~\eqref{eq:H-EM} with appropriate free-evolution, viscous damping terms $\delta\dot{\hat{\phi}}=-\gamma_{\text{tx}}\delta\hat{\phi}+\ldots$ and $\delta\dot{\hat{p}}=-(\gamma_{\text{m}}+\gamma_{\text{wg}})\delta\hat{p}+\ldots$, and associated noise/signal inputs $\hat{b}_{\text{in}}$, $\hat{a}_{\text{in,m}}$, and $\hat{a}_{\text{in}}$ yielding
\begin{eqnarray}
\dot{\hat{a}} & = & -i\bar{\omega}_{\text{m}}\hat{a}-\frac{\gamma_{\text{m}}+\gamma_{\text{wg}}}{2}(\hat{a}-\hat{a}^{\dagger})-i\frac{g}{2}(\hat{b}e^{i\omega_{\text{d}}t}+\hat{b}^{\dagger}e^{-i\omega_{\text{d}}t})\nonumber\\
&&-i\sqrt{\gamma_{\text{m}}}(\hat{a}_{\text{in,m}}+\hat{a}^{\dagger}_{\text{in,m}})-i\sqrt{\gamma_{\text{wg}}}(\hat{a}_{\text{in}}+\hat{a}^{\dagger}_{\text{in}}),\nonumber\\
\dot{\hat{b}} & = & -i\bar{\omega}_{\text{LC}}\hat{b}-\frac{\gamma_{\text{tx}}}{2}(\hat{b}-\hat{b}^{\dagger})-i\frac{g}{2}e^{-i\omega_{\text{d}}t}(\hat{a}+\hat{a}^{\dagger})\nonumber\\
&&-i\sqrt{\gamma_{\text{tx}}}(\hat{b}_{\text{in}}+\hat{b}^{\dagger}_{\text{in}}),\label{eq:example-EOM}
\end{eqnarray}
where $\gamma_{\text{tx}},\gamma_{\text{wg}}$ are the coupling rates to the transmission line and the phononic waveguide, respectively, while $\gamma_{\text{m}}$ represents unwanted, incoherent mechanical damping. 
The scattering dynamics of the transducer~\eqref{eq:scat-matrix-gen} can be determined by combining Eqs.~\eqref{eq:example-EOM} with the input-output relations
\begin{eqnarray}
\hat{\tilde{b}}_{\text{out}}(\Omega)&=&i\sqrt{\gamma_{\text{tx}}}\hat{\tilde{b}}(\Omega)+\hat{\tilde{b}}_{\text{in}}(\Omega),\nonumber\\
\hat{a}_{\text{out}}(\Omega)&=&i\sqrt{\gamma_{\text{wg}}}\hat{a}(\Omega)+\hat{a}_{\text{in}}(\Omega),\label{eq:example-IO}
\end{eqnarray}
valid for narrowband signals $||\Omega|-\bar{\omega}_{\text{m}}|\ll \bar{\omega}_{\text{m}}$; here we have introduced the rotating-frame electrical operators $\hat{\tilde{b}}(t)\equiv e^{i \omega_{\text{d}}t} \hat{b}(t)$ and $\hat{\tilde{b}}_{\text{in}}(t)\equiv e^{i \omega_{\text{d}}t} \hat{b}_{\text{in}}(t)$. The frequency-domain input and output field operators obey commutation relations $[\hat{\tilde{b}}_{\text{in/out}}(\Omega),\hat{\tilde{b}}^{\dagger}_{\text{in/out}}(\Omega')]=\delta(\Omega-\Omega')$ with analogous expressions for $\hat{a}_{\text{in/out}},\hat{a}_{\text{m,in/out}}$.
The present system exemplifies a more general family of transducers for which an explicit solution for $\mathbf{S}(\Omega)$ is provided in Appendix~\ref{sec:app-scat}. Considering for specificity electrical-to-mechanical transduction of electrical signals in the upper sideband, i.e., spectral components around $\Omega\sim\bar{\omega}_{\text{m}}$, the relevant scattering relation~\eqref{eq:scat-rel} yielding the mechanical readout of the downconverted electrical signal is ($\Omega >0$)
\begin{equation}
\hat{a}_{\text{out}}(\Omega) = U_{\text{s}}(\Omega)\hat{\tilde{b}}_{\text{in}}(\Omega) + \hat{\mathcal{F}}(\Omega),\label{eq:example-scat-rel}
\end{equation}
where the contribution from the lower electrical sideband $\hat{\tilde{b}}^{\dagger}_{\text{in}}(-\Omega)$ (assumed to contain no signal) is included in the noise operator $\hat{\mathcal{F}}(\Omega)$. The signal transfer function $U_{\text{s}}(\Omega)$ and the operator $\hat{\mathcal{F}}(\Omega)$ can be read off from the following relation, which represents an explicit example of (a particular row of) $\mathbf{S}(\Omega,\Omega')$,
\begin{multline}
\hat{a}_{\text{out}}(\Omega)=\hat{a}_{\text{in}}(\Omega)\\+i\sqrt{2\gamma_{\text{wg}}}
\chi_{\text{m}}(\Omega)
\Big(\sqrt{2\gamma_{\text{wg}}}\hat{a}_{\text{in}}(\Omega)+\sqrt{2\gamma_{\text{m}}}\hat{a}_{\text{in,m}}(\Omega)\\
-g\sqrt{2\gamma_{\text{tx}}}[\chi_{\text{LC,}+}(\Omega)\hat{\tilde{b}}_{\text{in}}(\Omega)
+\chi_{\text{LC,}-}(\Omega)\hat{\tilde{b}}_{\text{in}}^{\dagger}(-\Omega)]\Big),\label{eq:example-scat-rel2}
\end{multline}
expressed in terms of the effective mechanical susceptibility,
\begin{equation}
\chi_{\text{m}}(\Omega)\equiv[\chi_{\text{m,0}}^{-1}(\Omega)-g^{2}\sum_{s=\pm}\chi_{\text{LC,}s}(\Omega)]^{-1},\label{eq:chi-m}
\end{equation}
with the bare mechanical susceptibility being
\begin{equation}
\chi_{\text{m},0}(\Omega)=\frac{\bar{\omega}_{\text{m}}}{\bar{\omega}_{\text{m}}^{2}-\Omega^{2}- i\Omega(\gamma_{\text{m}}+\gamma_{\text{wg}})},
\end{equation}
and the rotating-frame circuit susceptibility to upper and lower sidebands ($\pm$) with respect to the drive frequency $\omega_{\text{d}}$,
\begin{equation}
\chi_{\text{LC,}\pm}(\Omega)=\frac{\bar{\omega}_{\text{LC}}}{\bar{\omega}_{\text{LC}}^{2}-(\omega_{{\rm d}}\pm\Omega)^{2}\mp i(\omega_{{\rm d}}\pm\Omega)\gamma_{\text{tx}}}.\label{eq:chi-LC}
\end{equation}
The ideal transduction scenario of 1-to-1 conversion of frequency components is possible in the resolved-sideband regime, i.e., $\chi_{\text{LC,}-}/\chi_{\text{LC,}+} \rightarrow 0$ for frequencies of interest, in which the lower electrical sideband, residing at lab frame frequencies $\omega \sim -\bar{\omega}_{\text{m}}+\omega_{\text{d}}$, is suppressed due to the circuit resonance being narrow compared to the sideband separation, $\gamma_{\text{tx}}/2\ll 2\bar{\omega}_{\text{m}}$.

\section{Figures of merit}
Having defined the generic family of transducers to be considered, Eqs.~(\ref{eq:scat-rel},\ref{eq:F-noise_def2}), we now proceed to identify suitable figures of merit.
Eq.~(\ref{eq:scat-rel}) shows that if we
consider the upper (lower) output sideband alone, $\Omega>0$ ($\Omega<0$),
the transducer is phase-preserving (phase-conjugating)~\cite{Caves1982}.
This prompts us to introduce the {\it signal transfer efficiency} for the upper
and lower sidebands as
\begin{equation}
\eta(\Omega) \equiv 
\begin{cases}
|U_{\text{s}}(\Omega)|^{2} & \text{for }\Omega>0\\
|V_{\text{s}}(\Omega)|^{2} & \text{for }\Omega<0\\
\end{cases}.\label{eq:eta-def}
\end{equation}
In the case where $V_{m}=0$ for all $m$, a transducer described by Eq.~(\ref{eq:scat-rel}) can be understood by the simple beam splitter model in Fig.~\ref{fig:Beam-splitter_RSB-transducer}b.
 In the ideal limit, a transducer losslessly converts photons from one frequency to another. In reality, however, photons may not be converted with unit efficiency $\eta<1$ and this loss is complemented by the admixture of noise $\hat{\mathcal{F}}$ from the other port 
 representing a superposition of noise sources. 

To illustrate Eq.~\eqref{eq:eta-def} we apply it to the example transducer of Section~\ref{sec:example}. Absorbing the dynamical ``electrical spring'' shift of the mechanical resonance implied by the effective susceptibility~\eqref{eq:chi-m} into $\bar{\omega}_{\text{m}}$ for simplicity and assuming $\omega_{\text{d}}=\bar{\omega}_{\text{LC}}-\bar{\omega}_{\text{m}}$, we find from Eqs.~(\ref{eq:example-scat-rel},\ref{eq:example-scat-rel2},\ref{eq:eta-def}) that the peak transfer efficiency is
\begin{equation}
\eta (\bar{\omega}_{\text{m}}) \approx\frac{4\gamma_{\text{wg}}g^{2}/\gamma_{\text{tx}}}{\left(\gamma_{\text{m}}+\gamma_{\text{wg}}+(g^{2}/\gamma_{\text{tx}})\left[1-\frac{(\gamma_{\text{tx}}/2)^2}{(2\bar{\omega}_{\text{m}})^2+(\gamma_{\text{tx}}/2)^2}\right]\right)^2},\label{eq:example-eta}
\end{equation}
approximating $\chi_{\text{LC},-}(\bar{\omega}_{\text{m}})$~\eqref{eq:chi-LC} by a Lorentzian for simplicity as is warranted when $\gamma_{\text{tx}},2\bar{\omega}_{\text{m}}\ll \bar{\omega}_{\text{LC}}$; Eq.~\eqref{eq:example-eta} is only meaningful for $\Omega >0$ seeing as the mechanical oscillator is described in the lab frame.
Ideal transduction requires $\eta \rightarrow 1$ and arises in the limit of negligible intrinsic transducer damping $\gamma_{\text{m}} \ll \gamma_{\text{wg}}$, resolved electrical sidebands $\gamma_{\text{tx}}/2\ll 2\bar{\omega}_{\text{m}}$, and impedance matching of the phononic waveguide to the induced electro-mechanical coupling rate $\gamma_{\text{wg}}=g^{2}/\gamma_{\text{tx}}$.

The quantity $\eta$ by itself is insufficient to characterize a transducer, as it says nothing about the transducer noise. A useful measure of the added noise can be obtained from Eqs.~(\ref{eq:scat-rel},\ref{eq:F-noise_def2}). Suppose that we are interested in measuring the total number of photons at the output during a time $T$.
It will have two contributions $\int (d\Omega/2\pi) \eta(\Omega) (\langle  \hat{a}_{{\rm in,s}}^\dagger  \hat{a}_{{\rm in,s}}\rangle(\Omega)+N(\Omega))$, where $\langle  \hat{a}_{{\rm in,s}}^\dagger  \hat{a}_{{\rm in,s}}\rangle(\Omega) \delta (\Omega - \Omega') \equiv \langle  \hat{a}_{{\rm in,s}}^\dagger (\Omega) \hat{a}_{{\rm in,s}}(\Omega')\rangle$ and
\begin{equation}
N(\Omega)\delta(\Omega-\Omega')=\frac{\langle \hat {\mathcal{F}}^\dagger(\Omega)\hat {\mathcal{F}}(\Omega')\rangle}{\eta (\Omega)}\label{eq:N-def-simple}
\end{equation}
quantifies the {\it added noise}. For measuring the input signal we are interested in knowing the output signal relative to the noise. This is exactly what is described by $N(\Omega)$, which signifies how many photons an input signal should have per mode in order to exceed the noise, i.e., the added noise flux per unit bandwidth referenced to the input. $N$ is thus the central quantity of interest in this case, and in particular $N\lesssim 1$ is desired for applications in the quantum regime, where we are sensitive to single photons. $N$ is closely related to the ambient temperature of the transducer, as can be seen by evaluating $\langle\hat{\mathcal{F}}^{\dagger}(\Omega)\hat{\mathcal{F}}(\Omega')\rangle$ under the assumption of time-stationary thermal
reservoirs $\langle \hat a_m^\dagger(\Omega) \hat a_m(\Omega')\rangle= n_m(\Omega+\omega_{0,m})\delta(\Omega-\Omega')$; here the mean number of thermal excitations for band $m$ is $n_m(\omega)=(\exp[\hbar\omega/k_{\text{B}}T_m]-1)^{-1} $, allowing for individual ambient temperatures $T_m$. From this we find that (considering $\Omega>0$ for specificity, the case $\Omega<0$ is similar)
\begin{multline}
\langle\hat{\mathcal{F}}^{\dagger}(\Omega)\hat{\mathcal{F}}(\Omega')\rangle=\delta(\Omega-\Omega') (\sum_{m\neq \mathrm{s}}
|U_{m}(\Omega)|^{2}n_{m}(\Omega+\omega_{0,m})\\+\sum_{m}|V_{m}(\Omega)|^{2}[n_{m}(-\Omega+\omega_{0,m})+1]),\label{eq:FF_exp-val}
\end{multline}
where the second sum includes the noise due to the coupling to the
lower sideband of the input port.
Knowing the efficiency $\eta$ from Eq.~(\ref{eq:eta-def}), we can use Eq.~(\ref{eq:FF_exp-val}) to evaluate the added noise $N$ (\ref{eq:N-def-simple}). 

Returning once again to the example system introduced in Section~\ref{sec:example}, we evaluate the added noise~\eqref{eq:N-def-simple} under the same assumptions used in Eq.~\eqref{eq:FF_exp-val}. At the mechanical resonance we arrive at
\begin{multline}
N(\bar{\omega}_{\text{m}}) \approx \frac{(\gamma_{\text{tx}}/2)^2}{(2\bar{\omega}_{\text{m}})^2+(\gamma_{\text{tx}}/2)^2}[n_{\text{tx}}(\bar{\omega}_{\text{LC}}-2\bar{\omega}_{\text{m}})+1]\\
\frac{\gamma_{\text{m}}}{g^2/\gamma_{\text{tx}}}n_{\text{m}}(\bar{\omega}_{\text{m}})+\left(\sqrt{\frac{\gamma_{\text{wg}}}{g^2/\gamma_{\text{tx}}}}-\frac{1}{\sqrt{\eta(\bar{\omega}_{\text{m}})}}\right)^{2} n_{\text{wg}}(\bar{\omega}_{\text{m}}),\label{eq:example-N}
\end{multline}
in the Lorentzian approximation as in Eq.~\eqref{eq:example-eta}. The ideal transduction limit of Eq.~\eqref{eq:example-N}, $N\rightarrow 0$, yields parameter criteria that are compatible with, but typically stricter than, those required for $\eta \rightarrow 1$ discussed below Eq.~\eqref{eq:example-eta}. Considering the first term in Eq.~\eqref{eq:example-N}, we see that the electrical sideband resolution must ensure $(\gamma_{\text{tx}}/2)\sqrt{n_{\text{tx}}(\bar{\omega}_{\text{LC}}-2\bar{\omega}_{\text{m}})+1}\ll 2\bar{\omega}_{\text{m}}$. Turning to the second term, this entails a requirement of large electro-mechanical quantum cooperativity, $g^2/(\gamma_{\text{tx}}\gamma_{\text{m}}n_{\text{m}}(\bar{\omega}_{\text{m}}))\gg 1$. The third term, representing the net reflected phononic waveguide noise, is suppressed due to impedance matching when $\gamma_{\text{wg}}=g^2/\gamma_{\text{tx}}$ and $1-\eta \ll 2/\sqrt{n_{\text{wg}}(\bar{\omega}_{\text{m}})}$.

The signal transfer efficiency $\eta$ and the added noise $N$, introduced in Eqs.~(\ref{eq:eta-def},\ref{eq:N-def-simple}), are the two essential parameters for characterizing the performance of a linear transducer (as we will demonstrate by considering various applications below).
Demanding that the transducer preserves the commutation relations of the itinerant fields~\cite{Caves1982}, Eqs.~(\ref{eq:scat-rel},\ref{eq:F-noise_def2},\ref{eq:eta-def}) imply
 ($\Omega>0$)
\begin{equation}
1=\eta(\Omega)+\sum_{m\neq\text{s}}|U_{m}(\Omega)|^{2}-\sum_{m}|V_{m}(\Omega)|^{2}.\label{eq:sum-rule}
\end{equation}
If we consider the limiting case where the scattering relation (\ref{eq:scat-rel}) describes a beam splitter interaction ($V_m=0$ for all $m$), corresponding to the fully resolved-sideband regime,
Eq.~(\ref{eq:sum-rule}) takes the form of a sum rule.
Hence if we achieve unit efficiency $\eta(\Omega)=1$, the contributions from all other noise sources vanish and we have an ideal transducer (for input signals centered at $\omega_{0,\mathrm{s}} + \Omega$). On the other hand if we consider imperfect transducers connecting vastly different frequency scales, we will have to appropriately balance the various noise contributions. Specifically, the thermal reservoir occupancies and corresponding `loads' on each channel
will be different, e.g., due to the vast gap between optical ($\gtrsim300\text{ THz}$),
mechanical and electrical frequencies ($\approx$1 MHz to 10 GHz). Thus, 
trade-offs have to be made between maximizing signal transfer
efficiency $\eta$ and minimizing the added noise $N$.
Outside the resolved-sideband regime ($V_{m} \neq 0$ for some $m$), Eq.~(\ref{eq:sum-rule}) 
allows $\eta(\Omega)>1$ at the cost of amplification noise. This further emphasizes
the above conclusion that maximization of $\eta(\Omega)$ in itself
is not a meaningful optimization strategy in general.


\section{Applications}
Having discussed the general features of transducers, we can evaluate their performance in terms of the parameters $\eta$ and $N$ for various applications.

\subsection{Optical heterodyne detection of upconverted signals}
As a particular application we first consider sensing of weak signals, e.g., upconverted electrical signal by optical means. We are interested in measuring  both quadratures of the incoming signal (phase-insensitive measurement) and will therefore consider heterodyne detection.  
Alternatively if only a single quadrature is desired or for quantum transduction and squeezing applications, homodyne detection may be advantageous.

\begin{figure}
\centering
\def\svgwidth{\columnwidth}
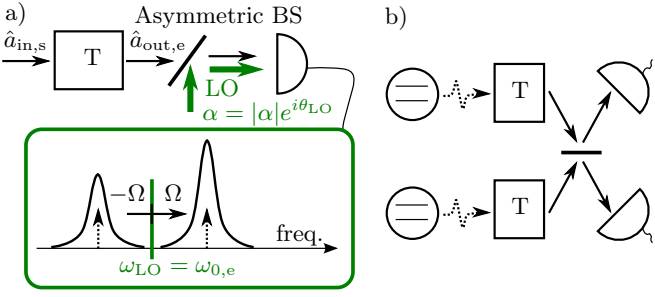
\caption{a) Heterodyne detection by mixing the output of the transducer (T) with a LO of frequency $\omega_{0,e}$ on a beam splitter. In general the Fourier component of the photocurrent $I(\Omega)$ at a frequency $\Omega$ will contain contributions from both the upper $\omega_{\text{LO}}+\Omega$ and lower sideband $\omega_{\text{LO}}-\Omega$. b) Entanglement generation by transducing the output from two qubits to optical frequencies and interfering the signals on a beam splitter.
A click in one of the detectors is an indication
that the single-click scheme has succeeded. Subsequently applying
a symmetric $\pi$-pulse to the qubits and conditioning on a second
click in a two-click scheme decreases the sensitivity to transduced
noise photons.
\label{fig:Homodyn-hetdyn-spectra}}
\end{figure}

Heterodyning relies on beating the transducer output
with a local oscillator (LO) of amplitude $\alpha_{\text{LO}}=|\alpha_{\text{LO}}|e^{i\theta_{\text{LO}}}$ at a well-defined frequency which we take to be at the center of the band
$\omega_{0,\mathrm{e}}$. With this choice the LO lies in between the two sidebands $\omega_{0,\mathrm{e}}\pm\Omega$ carrying the information to be measured.
Note that we consider the distinction of hetero- and homodyne detection from the perspective of the signal input mode, not the output mode. In particular, the considered strategy is a heterodyne detection of the input, but can be seen as a homodyne measurement of the output.
For simplicity we  confine
our attention to setups involving a single photo-detector as shown in Fig.~\ref{fig:Homodyn-hetdyn-spectra}a, where the LO is introduced via a highly asymmetric beam splitter. 
At the detector, 
the associated photocurrent is given by 
\begin{equation}
\hat{I}(\Omega)\approx\alpha_{\text{LO}}^{*}\hat{a}_{\text{out,e}}(\Omega)+\alpha_{\text{LO}}\hat{a}_{\text{out,e}}^{\dagger}(-\Omega)\ .\label{eq:homodyne-photo-current_main-text}
\end{equation}
The LO phase  determines the relative phase with which the sidebands enter the
linear combination.

Substituting the scattering relation of Eq.~(\ref{eq:scat-rel}) into Eq.~(\ref{eq:homodyne-photo-current_main-text}), we see that the spectral component $\hat{I}(\Omega)$ is phase preserving and directly proportional to the input we want to measure $\hat{I}(\Omega)\propto\hat a_{\mathrm{in,s}}(\Omega)$. Moreover, assuming an input coherent state $\langle \hat a_{\mathrm{in,s}}\rangle=\alpha$, the signal to noise ratio is phase independent and is given by $\delta(\Omega-\Omega')|\langle \hat{I}_\alpha(\Omega)\rangle|^2/\langle \hat{I}_{\alpha=0}(\Omega) \hat{I}_{\alpha=0}(\Omega')\rangle=|\alpha|^2/P_{\mathrm{s}}$. Here $\hat{I}_\alpha$ is the current with an incoming coherent state $\alpha$ and the power spectral noise density relative to the signal is given by 
\begin{multline}
P_{\mathrm{s}}(\Omega)=\frac{1}{2}+ \frac{1}{|t_{\theta_{\text{LO}}}(\Omega)|^{2}}\bigg[\eta(\Omega)N(\Omega)+\eta(-\Omega)N(-\Omega)\\+\frac{1}{2}+\frac{1-\eta(\Omega)+\eta(-\Omega)}{2}+\text{Re}\left[e^{-2i\theta_{\text{LO}}}f(\Omega)\right]\bigg]\ .
\label{eq:Sensitivity-peak_def}
\end{multline}
Here  the effective value of the transfer function is
 \begin{gather}
t_{\theta_{\text{LO}}}(\Omega)\equiv e^{-i\theta_{\text{LO}}}U_{\text{s}}(\Omega)+e^{i\theta_{\text{LO}}}V_{\text{s}}^{*}(-\Omega),
\end{gather}
and
\begin{gather}
f(\Omega)\delta(\Omega-\Omega')\equiv\langle\hat{\mathcal{F}}(\Omega)\hat{\mathcal{F}}(-\Omega')\rangle+\langle\hat{\mathcal{F}}(-\Omega)\hat{\mathcal{F}}(\Omega')\rangle
\end{gather}
is an  interference between the sidebands.
We note that since the two sidebands of the output arise from the same input fields, we may bound their mutual interference from above using the Cauchy-Schwarz inequality; choosing the LO phase so as to have constructive interference for the signal we arrive at (see Appendix~\ref{sec:Appendix-CS} for details)
\begin{multline}
P_s(\Omega)\leq\frac{1}{2}+
\bigg(w_{+}\sqrt{N(\Omega)+\frac{1}{2}[\frac{1}{\eta(\Omega)}-1]} \\
+w_{-}\sqrt{N(-\Omega)+\frac{1}{2}[\frac{1}{\eta(-\Omega)}+1]}\bigg)^{2},
\label{eq:P-peak_bound}\end{multline}
where $w_{\pm}\equiv \sqrt{\eta(\pm\Omega)}/(\sqrt{\eta(\Omega)}+\sqrt{\eta(-\Omega)})$.
From here we see that the sensitivity is primarily determined by a suitably weighted added noise whereas the efficiency mainly enters into the vacuum noise contribution (the last term inside the square roots). 

\subsection{Deterministic qubit transduction}
Next, as an application  for quantum information processing, we consider deterministic transduction of a qubit state.
For simplicity, we only consider
the upper-sideband output of the transducer although a better performance
might in principle be achieved by also including the lower sideband.
For a spectrally narrow input pulse with center frequency $\Omega_{\text{sig}}$ at the upper sideband $\hat{a}_{\text{s}} = \int h_{\text{in}}(\Omega)\hat{a}_{\text{in,s}}(\Omega)d\Omega$ (suitably normalized)
in the qubit state $|\psi\rangle_{\text{in}} = [\cos(\theta/2)+\sin(\theta/2)e^{i\phi}\hat{a}_{\text{s}}^{\dagger}]|0\rangle_{\text{in}}$, where $|0\rangle_{\text{in}}$ is the incoming vacuum of the input port, we find that the output state~$\hat{\rho}_{\text{out}}$ has the same noise in all quadratures and a resulting fidelity of (averaging over the Bloch sphere)~\cite{Sherson2006}
\begin{multline}
F_{\text{q}} =\frac{1}{4\pi}\int d\Theta \,\text{Tr}[\hat{\rho}_{\text{out}}|\psi\rangle_{\text{out}}\langle\psi|]\\
\approx 1 - \frac{5}{3}\eta^{(+)}N^{(+)} + \frac{2}{3}(\sqrt{\eta^{(+)}}-1) + \frac{1}{6}(\sqrt{\eta^{(+)}}-1)^2,
\label{eq:F-q_result_main-text}
\end{multline}
where we have defined $\eta^{(+)}\equiv \eta(\Omega_{\text{sig}})$ and $N^{(+)}\equiv N(\Omega_{\text{sig}})$, and we work in the limit $|\sqrt{\eta^{(+)}}-1| \ll 1,\eta^{(+)}N^{(+)} \ll 1$ ($|\psi\rangle_{\text{out}}$ is defined analogously to $|\psi\rangle_{\text{in}}$ in terms of $\hat{a}_{\text{out,e}}(\Omega)$). Hence $N$ and $\eta$
are again the crucial parameters for describing how well the transducer
performs.
Note that $\eta^{(+)} > 1$ in Eq.~(\ref{eq:F-q_result_main-text}) will inevitably be accompanied by amplification noise in $N^{(+)}$ due to Eq.~\eqref{eq:sum-rule}.

\subsection{Photon counting}
Finally, we  turn to discrete variable  photon counting of the output signal.
The role of the transducer in this case is to perform
frequency conversion of each photon. To this end, the beam splitter interaction ($V_i\approx 0$) is
desirable since it directly converts quanta from one frequency to
another. 
We shall therefore
consider transducers which are reasonably sideband-resolved. Nevertheless non-zero temperature as well as imperfect sideband resolution will still lead to photons leaving the transducer giving rise to an effective dark count rate.  
As opposed to the heterodyne measurement considered above, photon counting is not mode-selective and will count photons
of all modes impinging on the detector. Using the scattering relation in Eq.~(\ref{eq:scat-rel}) the dark count rate can be expressed as
\begin{equation}
r_{\rm N}=\eta^{(+)}N^{(+)} B,\label{eq:r-N}
\end{equation}
where we have separated out the efficiency at the signal peak $\eta^{(+)}$ and the corresponding added noise $N^{(+)}$ and introduced 
\begin{equation}
B=\int \frac{d\Omega}{2\pi} \frac{\eta(\Omega)}{\eta^{(+)}} \frac{N(\Omega)}{N^{(+)}}. 
\end{equation}
If the added noise $N(\Omega)$ can be considered constant over the entire bandwidth of the transducer, we can interpret $B$ as a measure of the bandwidth.

We consider again an incoming temporal mode~$\hat{a}_{\text{s}}$ as introduced above Eq.~(\ref{eq:F-q_result_main-text}).
If we integrate over the entire output, the mode-dependent efficiency is given by
\begin{equation}
\eta_{h}=\int_{-\infty}^{\infty}\eta(\Omega)|h_{\text{in}}(\Omega)|^{2}d\Omega.\label{eq:eta-h}
\end{equation}
Introducing the normalized mode function $h_{\text{out}}(t)$ for the output and considering a single photon in the input, we may express the number of photons counted during a time interval $T$ as 
\begin{equation}
\bar{n}_{\text{out}}=\eta^{(+)}\left(\frac{\eta_{h}}{\eta^{(+)}}\int_{0}^{T}|h_{\text{out}}(t)|^{2}dt+N^{(+)}BT\right).\label{eq:discr-var_n_out-mean}
\end{equation}
Here the first term in the parenthesis  represents the desired component. This term is upper bounded by unity, which can only be reached in the limit of a very long time interval $T$. Hence the added noise  relative to the signal is again given by the added noise $N^{(+)}$, but now it is increased by a factor of $BT\gtrsim 1$ since photon counters are not mode selective. 

\subsection{Transducer-mediated conditional entanglement of remote qubits}
As an application for quantum information processing in the discrete variable regime, we consider the remote entanglement of two atom-like systems, e.g., superconducting qubits by transducing the signal to optical frequencies for long distance communication as shown in Fig.~\ref{fig:Homodyn-hetdyn-spectra}b.
This is of particular interest for (entanglement-based) quantum repeaters,
which may allow for the realization of a long-ranging quantum internet
based on an optical fiber infrastructure~\cite{Kimble2008}. To achieve this based on
super-conducting systems, transduction between microwave and optical
frequencies is required.

A number of different protocols have been suggested for conditional entanglement generation along these lines in the context of distant trapped atoms~\cite{Moehring2007}. In particular, protocols relying on a single click are advantageous for low $\eta\ll1$ since they give a higher success probability, whereas two click protocols are advantageous in terms of the resulting fidelity. Alternatively, one can condition on a continuous-variable measurement, as has previously been analyzed in the present context of transducer-mediated entanglement between qubits~\cite{Ondrej2016}.

The entanglement schemes considered here involve the emission of single photons from
the (artificial) atoms which need to be transduced to optical frequencies
for fiber transmission (see Fig.~\ref{fig:Homodyn-hetdyn-spectra}b). 
For simplicity, we will in this analysis neglect parametric amplification
effects by assuming it very unlikely that a single incoming signal
quantum generates more than one quantum at the exit port. 
This allows us to obtain the transduction efficiency for a single
signal quantum: 
\begin{equation}
\eta=\eta_{h}\int_{0}^{T}|h_{\text{out}}(t)|^{2}dt,\label{eq:eta-def_Appendix}
\end{equation}
with $\eta_{h}$ as defined in Eq.~\eqref{eq:eta-h}. The inevitable addition
of noise photons amounts to an additional equivalent dark count probability
$P_{\text{d}}$ related to the noise rate $r_{\text{N}}$~\eqref{eq:r-N}; here we shall take this to be the only source of
dark counts. 
To simplify the analysis we make the assumptions that the noise photons
are either distinguishable or that the overall efficiency is low, preventing bunching effects.
From this assumption it follows that each
transducer contributes an average dark count rate of $r_{\text{N}} T/2$
in each detector, whereby the probability for at least one dark count
in a particular detector is $P_{\text{d}}=1-(e^{-r_{\text{N}} T/2})^{2}=1-e^{-r_{\text{N}} T}\approx r_{\text{N}} T$
for $r_{\text{N}} T\ll1$.

The basic idea of the schemes to be considered here is to symmetrically
excite the artifical atoms into a state of the form \cite{Moehring2007}
\begin{multline}
(\sqrt{1-P_{\text{e}}}|0\rangle_{\text{A,}1}|0\rangle_{\text{P,}1}+\sqrt{P_{\text{e}}}|1\rangle_{\text{A,}1}|1\rangle_{\text{P,}1})\\
\otimes(\sqrt{1-P_{\text{e}}}|0\rangle_{\text{A,}2}|0\rangle_{\text{P,}2}+\sqrt{P_{\text{e}}}|1\rangle_{\text{A,}2}|1\rangle_{\text{P,}2}),\label{eq:discr-var_sym-atom-excitation}
\end{multline}
where $|n\rangle_{\text{A,}i}$ denotes the atomic state of atom $i$
and $|n\rangle_{\text{P,}i}$ the photonic Fock states corresponding
to the emitted light from atom $i$. The photonic states are upconverted to the optical domain
by individual transducers and mixed in a mode-matched
fashion at a 50:50 beamsplitter, thereby withholding the which-way
information from the subsequent photodetection measurement (see Fig.~\ref{fig:Homodyn-hetdyn-spectra}b). Hence, in absence of
imperfections, if $P_e \ll 1$ and a single click is obtained, the atomic system is
projected into an entangled state of either atom being in its $|1\rangle_{\text{A},i}$
state:
\begin{equation}
|\Psi_{\pm}\rangle=\frac{1}{\sqrt{2}}(|0\rangle_{\text{A,}1}|1\rangle_{\text{A,}2}\pm|1\rangle_{\text{A,}1}|0\rangle_{\text{A,}2}),\label{eq:Psi_pm-main_text}
\end{equation}
with the sign determined by which detector clicks. For higher efficiencies, multi-photon events would be problematic for this approach. In that case, a two-click scheme, which
adds a subsequent $\pi$-pulse along with the condition of an additional
click, serves to verify that the atomic systems are in the state
(\ref{eq:Psi_pm-main_text}). This added step mitigates the effect
of dark counts and atomic double
excitations, hence allowing $P_{\text{e}}=1/2$.

We now calculate the conditional fidelities $F_{i\text{c}}$, $i\in\{1,2\}$
for Bell-state generation by means of these single-click and two-click
variants.
This conditional fidelity is defined as the average overlap between
the generated and desired states given that the relevant click condition
was fulfilled. Upon fulfillment of the condition, the system is described
by a certain density matrix $\hat{\rho}_{i\text{c}}$. Starting with
the single-click condition, we will now determine the conditional
fidelity of achieving either of the states $|\Psi_{\pm}\rangle$.
This may be calculated by imagining that if we obtain $|\Psi_{-}\rangle$,
we rotate it into $|\Psi_{+}\rangle$; denoting the corresponding
rotated density matrix $\hat{\rho}_{i\text{c}}'$, the desired conditional
fidelity is given by:
\begin{equation}
F_{i\text{c}}=\text{Tr}[\hat{\rho}_{i\text{c}}'|\Psi_{+}\rangle\langle\Psi_{+}|].\label{eq:F-c_single-photon}
\end{equation}
By considering the various possible outcomes compatible with fulfillment
of the condition in the limit $P_{\text{d}}/\eta\ll P_{\text{e}}\ll1$,
we arrive at (using the abbreviated notation $|i\rangle_{\text{A,}1}|j\rangle_{\text{A,}2}\equiv|ij\rangle$)
\begin{widetext}
\begin{multline}
\hat{\rho}_{1\text{c}}'=\frac{1}{\mathcal{N}_{1}}\left[\overbrace{(1-P_{\text{e}})^{2}2P_{\text{d}}(1-P_{\text{d}})|00\rangle\langle00|}^{\text{Neither emits, dark count in one arm}}+\overbrace{2P_{\text{e}}(1-P_{\text{e}})\eta(1-P_{\text{d}})|\Psi_{+}\rangle\langle\Psi_{+}|}^{\text{One atom emits and is detected, no dark count in other arm}}\right.\\
+\overbrace{P_{\text{e}}(1-P_{\text{e}})(1-\eta)2P_{\text{d}}(1-P_{\text{d}})[|01\rangle\langle01|+|10\rangle\langle10|]}^{\text{One emits but is not detected, dark count in one arm}}\\
\left.+\overbrace{P_{\text{e}}^{2}([1-(1-\eta)^{2}]+(1-\eta)^{2}2P_{\text{d}})(1-P_{\text{d}})|11\rangle\langle11|}^{\text{Both emit, if detected no dark count in other arm}}\right],\label{eq:discr-var_rho-single-photon}
\end{multline}
\end{widetext}
where $\mathcal{N}_{1}$ is the normalization factor that ensures
$\text{Tr}[\hat{\rho}_{1\text{c}}']=1$. Using Eqs.~(\ref{eq:Psi_pm-main_text})
and (\ref{eq:discr-var_rho-single-photon}) to evaluate the conditional
fidelity (\ref{eq:F-c_single-photon}), we find
\begin{equation}
F_{1\text{c}}=\frac{2P_{\text{e}}(1-P_{\text{e}})\eta+P_{\text{e}}(1-P_{\text{e}})(1-\eta)2P_{\text{d}}}{P_{\text{e}}\eta(1-2P_{\text{d}})[2-P_{\text{e}}\eta]+2P_{\text{d}}}.\label{eq:F-c_single-photon_exact-result}
\end{equation}
Expanding Eq.~(\ref{eq:F-c_single-photon_exact-result}) in the limit
$P_{\text{e}},P_{\text{d}}\ll1$, $P_{\text{d}}\ll P_{\text{e}}$
it reduces to 
\begin{equation}
F_{1\text{c}}\approx1-P_{\text{e}}(1-\eta/2)-\frac{P_{\text{d}}}{\eta P_{\text{e}}}.\label{eq:F-c_single-photon_approx-result}
\end{equation}
The choice of $P_{\text{e}}$ that maximizes $F_{1\text{c}}$ as given
by Eq.~(\ref{eq:F-c_single-photon_approx-result}) is
\begin{equation}
P_{\text{e}}^{\text{(opt)}}=\sqrt{\frac{P_{\text{d}}}{\eta(1-\eta/2)}},
\end{equation}
yielding the fidelity
\begin{eqnarray}
F_{1\text{c}}^{\text{(opt)}}&=&1-2\sqrt{\left(\frac{1}{\eta}-\frac{1}{2}\right)P_{\text{d}}}\nonumber\\
&\approx&1-2\sqrt{\left(\frac{1}{\eta}-\frac{1}{2}\right)\eta^{(+)}N^{(+)}B T},
\label{eq:F-1c-opt}
\end{eqnarray}
with success probability $P_{\mathrm{S}} \approx 2 \eta P_{\mathrm{e}}$.

Next, we consider the two-click scheme. The scheme works in two steps
and we will take as our condition that at least one click in exactly
one arm occurs in each of the two steps. In the first step the two
atoms are excited symmetrically to the state (\ref{eq:discr-var_sym-atom-excitation})
and preferably
only one of the atoms emit a photon. In the next step, a $\pi$-pulse
is applied symmetrically to the two atoms such that each atom flips between 
$|0\rangle_{\text{A}}$ and $|1\rangle_{\text{A}}$ states. A subsequent photon cycling event 
causes the remaining atom in the $|1\rangle_{\text{A}}$ state to emit. Only the case with only one atom in the state $|1\rangle_{\text{A}}$ leads to a click in each round of the protocol. Thus, in the absence of dark counts
and for perfect transduction, $P_{\text{d}}=0$, $\eta=1$, fulfillment
of the two-click condition means that either of the entangled atomic
states $|\Psi_{\pm}\rangle$~(\ref{eq:Psi_pm-main_text}) have
been generated with unit conditional fidelity (whether the first click
occurs in detector one or two reveals which of the two states where
generated). For finite dark count probability $P_{\text{d}}$, the
conditional fidelity drops below unity according to an expression to be
determined shortly. Fulfillment of the two-click condition corresponds
to the density matrix (rotating $|\Psi_{-}\rangle$ into $|\Psi_{+}\rangle$
for purposes of calculating $F_{2\text{c}}$, $\hat{\rho}_{2\text{c}}\rightarrow\hat{\rho}_{2\text{c}}'$)
\begin{widetext}
\begin{multline}
\hat{\rho}_{2\text{c}}'=\frac{1}{\mathcal{N}_{2}}\left[(1-P_{\text{e}})^{2}2P_{\text{d}}(1-P_{\text{d}})\left([1-(1-\eta)^{2}](1-P_{\text{d}})+(1-\eta)^{2}2P_{\text{d}}(1-P_{\text{d}})\right)|00\rangle\langle00|\right.\\
+P_{\text{e}}(1-P_{\text{e}})\left[\left(\eta(1-P_{\text{d}})+(1-\eta)2P_{\text{d}}(1-P_{\text{d}})\right)^{2}-\eta^{2}(1-P_{\text{d}})^{2}\right][|01\rangle\langle01|+|10\rangle\langle10|]\\
\left.2P_{\text{e}}(1-P_{\text{e}})\eta^{2}(1-P_{\text{d}})^{2}|\Psi_{+}\rangle\langle\Psi_{+}|+P_{\text{e}}^{2}\left((1-\eta)^{2}2P_{\text{d}}+(1-(1-\eta)^{2})\right)2P_{\text{d}}(1-P_{\text{d}})^{2}|11\rangle\langle11|\right].
\end{multline}
\end{widetext}
From this the conditional fidelity for entanglement generation in
the two-photon scheme is, from Eq.~(\ref{eq:F-c_single-photon}),
(evaluating at the optimum excitation probability $P_{\text{e}}=1/2$)

\begin{equation}
F_{2\text{c}}^{\text{(opt)}}=\frac{2P_{\text{d}}^{2}(1-\eta)^{2}+2P_{\text{d}}(1-\eta)\eta+\eta^{2}}{8P_{\text{d}}^{2}(1-\eta)^{2}+2P_{\text{d}}(4-3\eta)\eta+\eta^{2}},\label{eq:F-c_two-photon_exact}
\end{equation}
with success probability $P_{\mathrm{S}} \approx \eta^2/2$.

Comparing the single-click and two-click schemes, we find in the limit
$P_{\text{d}},P_{\text{d}}/\eta\ll1,\eta^{(+)}N^{(+)}B T\ll1$
that
\begin{eqnarray}
F_{1\text{c}} & \approx & 1-2\sqrt{\left(\frac{1}{\eta}-\frac{1}{2}\right)\eta^{(+)}N^{(+)}B T}\nonumber \\
F_{2\text{c}} & \approx & 1-\left(\frac{6}{\eta}-4\right)\eta^{(+)}N^{(+)}B T.\label{eq:Cond_fid-1-2_Appendix}
\end{eqnarray}
From these expressions we
see that the dependence of the fidelities on the efficiency $\eta$
is rather weak in both cases: It serves to determine a prefactor to
$N^{(+)}$ varying by at most a factor of 2 for $F_{\text{1c}}$
and at most a factor of 3 for $F_{\text{2c}}$, where we can have $\eta\sim\eta^{(+)}$ for a rather long pulse $B T\gtrsim 1$.
For instance for the one-click scheme we have the prefactor $(1/\eta-1/2)\eta^{(+)}\sim \eta^{(+)}/\eta\sim 1$ for $\eta \sim \eta^{(+)} \ll 1$. On the other hand, for $\eta\sim\eta^{(+)}\sim 1$ we have the prefactor $1/2$. 
The fact that the fidelity mainly depends on $N^{(+)}$ reflects 
that the conditional fidelity is determined by the probability to
detect the good transduced photons relative to the noise photons,
which is exactly determined by the added noise $N^{(+)}$.
Taking the low-efficiency limit $\eta=\eta^{(+)}\rightarrow0$ in the expressions for $F_{i\text{c}}$ given in Eqs.~(\ref{eq:Cond_fid-1-2_Appendix}), we arrive at
\begin{eqnarray}
F_{1\text{c}} & \approx & 1-2\sqrt{N^{(+)}BT}\nonumber \\
F_{2\text{c}} & \approx & 1-6N^{(+)}BT\label{eq:Cond_fid-1-2_main-text}
\end{eqnarray}
for the one- and two-click protocols, respectively. Here we  have considered a situation corresponding to long distance communication  $\eta\ll 1$ and assumed $BT\gg 1$ so that the the pulse fits  within both the spectral and temporal windows. As is evident, the key quantity for the quality of the generated entanglement is the added noise, whereas the efficiency of the transducer only enters into the success probability.

From the expressions~(\ref{eq:Cond_fid-1-2_Appendix},\ref{eq:Cond_fid-1-2_main-text}) it is clear
that the two-photon scheme has a smaller sensitivity to added noise
than the one-photon scheme in the interesting regime $N^{(+)}B T\ll1$.
On the other hand the two-photon scheme will have a lower success
probability if the transducer has a low efficiency since it requires
the detection of two photons. If we are only interested in the quality
of the produced entanglement, $N^{(+)}$ is the important quantity
to consider. As opposed to the situation for heterodyne detection,
where a single mode was measured, there is, however, an additional
factor coming from the fact the photo-detectors are not mode selective.
Since efficient transduction requires $B T>1$ this factor puts
an additional requirement on the added noise for photo-detection
schemes compared to continuous-variable schemes. On the other
hand, photo-detection schemes can give useful output even with limited
efficiency.

\section{Conclusion}
We have given a generic characterization of time-stationary transducers in terms of signal transfer efficiency $\eta$ and added noise $N$. The non-equilibrium character of transduction requires trade-offs in optimizing these quantities. By deriving the figures of merit for various quantum optics applications in terms of $\eta$ and $N$, we have clarified the requirements on a transducer to perform efficiently in each of these contexts. The examples considered here show that the added noise $N$ often plays a more important role than the signal transfer efficiency $\eta$ in determining the performance.

\begin{acknowledgments}
We acknowledge helpful conversations with J.~Borregaard, E.~Polzik, K.~Usami, J.~Aumentado, K.~Lenhert, and A.~Clerk. JMT thanks the NBI group for their hospitality during his stays. Likewise, EZ thanks the JQI for hosting him.
The research leading to these results was funded by The European Union Seventh Framework Programme through SIQS (grant no. 600645), ERC Grants QIOS (grant no. 306576), and Q-CEOM (grant no. 638765), as well as the ARL CDQI. EZ acknowledges funding from the Carlsberg foundation.
\end{acknowledgments}

\appendix

\section{Scattering matrix for linearized two-body interactions}\label{sec:app-scat}

We present here a family of coupling schemes for which the scattering matrix in Eq.~(\ref{eq:scat-matrix-gen}) of the main text takes the particular form of Eqs.~(\ref{eq:scat-rel},\ref{eq:F-noise_def2}).
To be concrete, we first derive the input-output relation~\cite{Collett1985} for a general transducer based on two-body interactions and driven by harmonic bias fields of constant amplitude. 
Linearizing such an open system around its drive-induced steady state gives Heisenberg-Langevin equations of motion of the following form:
\begin{gather}
\dot{\vec{B}}(t)=\sum_{k,l} \mathbf{M}_{l,k}{\rm e}^{il\omega_{\text{d},k}t}\vec{B}(t)-\mathbf{\Gamma}\vec{A}_{\text{in}}(t)\label{eq:IO-formalism-Adot}.
\end{gather}
Here $\vec{B}$ is a vector of bosonic operators describing the internal degrees of the freedom of the transducer, whose mutual coupling are accounted for by the matrices~$\mathbf{M}_{l,k}$, as induced by the various harmonic driving fields of frequency~$\omega_{\text{d},k}$, which may enter to various order $l=0,\pm1,\pm2,\ldots$ (depending on the interactions underlying the linearized theory). Decay of the internal modes~$\vec{B}$ entails coupling to the input fields~$\vec{A}_{\text{in}}$, as described by the matrix~$\mathbf{\Gamma}$ in~(\ref{eq:IO-formalism-Adot}), and the output fields~$\vec{A}_{\text{out}}$ according to
\begin{equation}
\vec{A}_{\text{out}}(t)=\mathbf{\Gamma'}\vec{B}(t)+\vec{A}_{\text{in}}(t),\label{eq:IO-rel}
\end{equation}
for a suitable matrix $\mathbf{\Gamma'}$.

\begin{figure}
\centering
\includegraphics[width=.75\columnwidth]{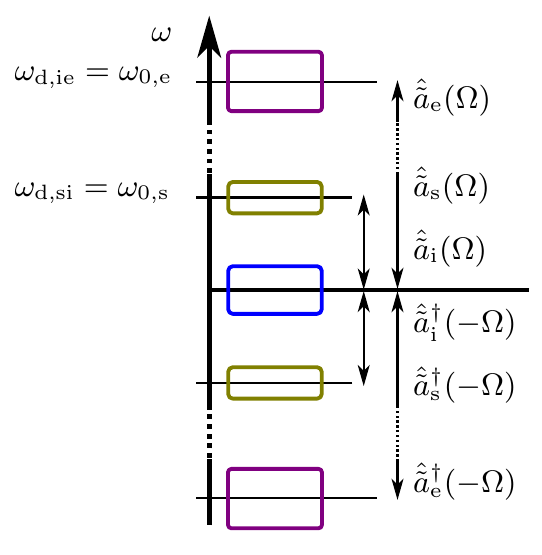}

\caption{Network topology of the internal modes of the transducer. The dynamics of the internal modes of the transducer are assumed to occur in narrow frequency bands centered around $\omega_{0,m}$. The harmonic driving terms $\omega_{\text{d},k}$ connect the different frequency bands, which can be at vastly different frequency scales (as indicated by dotted arrows and axis). Drive terms not matching the difference between the bands have been discarded in a rotating wave approximation. In this instance, the internal modes `s' and `e' that couple to the itinerant fields $\hat{a}_{\text{in,s}}$ and $\hat{a}_{\text{out,e}}$, respectively, are linked via the internal transducer mode `i', which has a central frequency $\omega_{0,\mathrm{i}}=0$.\label{fig:Freq-bands}}
\end{figure}

To proceed, we assume that the different field operators can be divided into a finite number of (narrow) frequency bands, each denoted by an index $m$ and centered around a frequency $\omega_{0,m}$ (possibly zero). We choose these central frequencies such that all bands connected by a non-zero matrix element of $\mathbf{M}_{l,k}$ in~(\ref{eq:IO-formalism-Adot}) are related to one another by an integer number $l$ of drive frequencies~$\omega_{\text{d,}k}$. Crucially, we make the assumption that the frequency bands are well-separated for all bands coupled by a time-varying field, $\omega_{\text{d,}k'}>0$, whereas we assume bands coupled due to a DC-bias, $\omega_{\text{d,}k'}=0$, to be overlapping (and have a common center frequency $\omega_{0,m'}$). 
These assumptions provide a unique rotating frame for our many-mode system, allowing removal of all time dependence in Eq.~(\ref{eq:IO-formalism-Adot}), and define a network topology in frequency space.
Furthermore, we make a rotating wave-type approximation by neglecting all matrix elements of $\mathbf{M}_{l,k}$ which are not connecting frequency bands.
Such a scenario is depicted in Fig.~\ref{fig:Freq-bands} and refers to the typical scenario where bias fields are used to frequency-convert between different components of the spectra.
After introducing the Fourier representation of all operators in the equation of motion (\ref{eq:IO-formalism-Adot}),   it is  convenient to change to a rotating frame where all field annihilation operators  for a particular band $\hat a(\Omega+\omega_{0,m})$ are replaced by slowly varying operators $\hat {\tilde a}(\Omega)=\hat a(\Omega+\omega_{0,m})$ with commutation relation $[\hat {\tilde a}(\Omega), \hat {\tilde a}^\dagger(\Omega')]=\delta(\Omega-\Omega')$. The resulting equations of motion only involve the slow frequency component $\Omega$. Solving Eq.~(\ref{eq:IO-formalism-Adot}) in the Fourier domain and using Eq.~(\ref{eq:IO-rel}) we finally find a scattering relation of the form given in Eq.~(\ref{eq:scat-matrix-gen}) with the scattering matrix (in the rotating frame)
\begin{gather}
\mathbf{S}(\Omega,\Omega')\equiv{\left(\mathbf{1}+\mathbf{\Gamma'}\frac{1}{i\Omega\mathbf{1}+\mathbf{M}}\mathbf{\Gamma}\right)}\delta(\Omega-\Omega').
\end{gather}
Here $\mathbf{M}=\sum_{\langle l,k \rangle }\mathbf{\tilde M}_{l,k}$ with the sum over terms $\langle l,k \rangle $ contributing within the rotating wave approximation, and the tilde on $\mathbf{\tilde M}_{l,k}$ denotes that terms corresponding to the central frequencies $\omega_{0,m}$ have been removed.  Note, that since annihilation operators enter with the time dependence  $\hat {\tilde a}(\Omega)\exp(-i\Omega t )$ whereas the creation operators are $\hat {\tilde a}^\dagger(\Omega)\exp(i\Omega t )$ the annihilation operators $\hat {\tilde a}(\Omega)$ will in general couple to $\hat {\tilde a}^\dagger(-\Omega)$ and the input (output) vectors $\vec{A}_{\text{in}}(\Omega)$ [$\vec{A}_{\text{out}}(\Omega)$] thus contain $\hat {\tilde a}(\Omega)$ and $\hat {\tilde a}^\dagger(-\Omega)$, which amounts to a folding of the input spectra onto themselves around the band center frequency~$\omega_{0,m}$ (with some gain profile). For simplicity we will in the main text primarily deal with the slowly varying operators and omit the tilde.

\section{Cauchy-Schwarz upper bound for heterodyne sensitivity}\label{sec:Appendix-CS}

Combining Eqs.~(\ref{eq:scat-rel}) and (\ref{eq:homodyne-photo-current_main-text})
from the main text we find that the Fourier transformed heterodyne
current has the following signal and noise components 
\begin{equation}
\hat{I}(\Omega)/|\alpha_{\text{LO}}|=t_{\text{s},\theta_{\text{LO}}}(\Omega)\delta\hat{a}_{\text{in,s}}(|\Omega|)+\hat{\mathcal{N}}_{\theta_{\text{LO}}}(\Omega),
\end{equation}
where we have defined
\begin{align}
t_{\text{s},\theta_{\text{LO}}}(\Omega) &\equiv e^{-i\theta_{\text{LO}}}U_{\text{s}}(\Omega)+e^{i\theta_{\text{LO}}}V_{\text{s}}^{*}(-\Omega),\\
\hat{\mathcal{N}}_{\theta_{\text{LO}}}(\Omega) &\equiv e^{-i\theta_{\text{LO}}}\hat{\mathcal{F}}(\Omega)+e^{i\theta_{\text{LO}}}\hat{\mathcal{F}}^{\dagger}(-\Omega).\label{eq:N-noise-operator_def}
\end{align}
Integrating the photocurrent with a cosine with a variable phase $\phi$, we see
that all input quadratures are contained in $\hat{I}(\Omega)$ obtained
for a fixed value of $\theta_{\text{LO}}$ 
\begin{gather}
\hat{Z}_{\phi,\theta_{\text{LO}}}(\Omega)  \equiv  \frac{1}{|\alpha_{\text{LO}}|}\int\hat{I}(t)\cos(\omega t+\phi)dt\nonumber\\
=\frac{e^{i\phi}\hat{I}(\Omega)+e^{-i\phi}\hat{I}^{\dagger}(\Omega)}{2|\alpha_{\text{LO}}|} \nonumber\\
  =  \frac{1}{\sqrt{2}}\left[|t_{\text{s},\theta_{\text{LO}}}(\Omega)|\hat{X}_{\text{s,}-(\psi+\phi)}(\Omega)+\hat{Y}_{\text{n}}(\Omega)\right],\label{eq:quad-scat-rel}
\end{gather}
where we have introduced the phase $\psi\equiv\text{Arg}[t_{\text{s},\theta_{\text{LO}}}(\Omega)]$
of the quadrature transfer function (\ref{eq:N-noise-operator_def})
of the single-mode input signal quadratures
\begin{equation}
\hat{X}_{\text{s,}\varphi}(\Omega)\equiv\frac{e^{-i\varphi}\delta\hat{a}_{\text{in,s}}(\Omega)+e^{i\varphi}\delta\hat{a}_{\text{in,s}}^{\dagger}(\Omega)}{\sqrt{2}},\label{eq:X-s_def}
\end{equation}
obeying the canonical commutation relations $[\hat{X}_{\varphi}(\Omega),\hat{X}_{\varphi+i\pi/2}(\Omega')]=i\delta(\Omega-\Omega')$;
the added quadrature noise in Eq.~(\ref{eq:quad-scat-rel}) is accounted
for by the Hermitian operator
\begin{equation}
\hat{Y}_{\text{n}}(\Omega)\equiv\frac{e^{i\phi}\hat{\mathcal{N}}_{\theta_{\text{LO}}}(\Omega)+e^{-i\phi}\hat{\mathcal{N}}_{\theta_{\text{LO}}}^{\dagger}(\Omega)}{\sqrt{2}},\label{eq:Y-noise-operator_def}
\end{equation}
where $\hat{\mathcal{N}}_{\theta_{\text{LO}}}$ was defined in Eq.~(\ref{eq:N-noise-operator_def}). 

Referencing Eq.~(\ref{eq:quad-scat-rel}) to the input signal, we
define the heterodyne sensitivity as the variance
\begin{widetext}
\begin{equation}
P(\Omega)\delta(\Omega-\Omega')\equiv\left\langle \left(\hat{X}_{\text{s,}-(\psi+\phi)}(\Omega)+\frac{\hat{Y}_{\text{n}}(\Omega)}{|t_{\text{s},\theta_{\text{LO}}}(\Omega)|}\right)\left(\hat{X}_{\text{s,}-(\psi+\phi)}(\Omega')+\frac{\hat{Y}_{\text{n}}(\Omega')}{|t_{\text{s},\theta_{\text{LO}}}(\Omega')|}\right)\right\rangle _{\text{vac,s}}
=\frac{1}{2}\delta(\Omega-\Omega')+\frac{\langle\hat{Y}_{\text{n}}(\Omega)\hat{Y}_{\text{n}}(\Omega')\rangle}{|t_{\text{s},\theta_{\text{LO}}}(\Omega)|^{2}},\label{eq:P-def_Appendix}
\end{equation}
\end{widetext}
where we take the input on the signal port to be vacuum which is uncorrelated
with the noise inputs. To evaluate $\langle\hat{Y}_{\text{n}}(\Omega)\hat{Y}_{\text{n}}(\Omega')\rangle$
we will make use of the property that the noise associated with $\hat{\mathcal{F}}$
is time-stationary, i.e.
\begin{gather}
\langle\hat{\mathcal{F}}^{\dagger}(\Omega)\hat{\mathcal{F}}(\Omega')\rangle\propto\delta(\Omega-\Omega'),\;\langle\hat{\mathcal{F}}(\Omega)\hat{\mathcal{F}}^{\dagger}(\Omega')\rangle\propto\delta(\Omega-\Omega')\nonumber \\
\langle\hat{\mathcal{F}}(\Omega)\hat{\mathcal{F}}(\Omega')\rangle\propto\delta(\Omega+\Omega'),\;\langle\hat{\mathcal{F}}^{\dagger}(\Omega)\hat{\mathcal{F}}^{\dagger}(\Omega')\rangle\propto\delta(\Omega+\Omega'),\label{eq:F-noise_exp-vals}
\end{gather}
which follows from the assumed form of $\hat{\mathcal{F}}$~(\ref{eq:F-noise_def2})
combined with the thermal expectation values
of the input operators $\hat{a}_{\text{in},i}(\Omega)$. In this way we find that
\begin{widetext}
\begin{eqnarray}
\langle\hat{Y}_{\text{n}}(\Omega)\hat{Y}_{\text{n}}(\Omega')\rangle &=&\frac{1}{2}\left[\langle\hat{\mathcal{N}}_{\theta_{\text{LO}}}(\Omega)\hat{\mathcal{N}}_{\theta_{\text{LO}}}^{\dagger}(\Omega')\rangle+\langle\hat{\mathcal{N}}_{\theta_{\text{LO}}}^{\dagger}(\Omega)\hat{\mathcal{N}}_{\theta_{\text{LO}}}(\Omega')\rangle\right]\nonumber\\
&=&\left|\left(\begin{array}{c}
\vec{u}^{(+)}\\
\vec{v}^{(+)}
\end{array}\right)\right|^{2}+\left|\left(\begin{array}{c}
\vec{v}^{(-)}\\
\vec{u}^{(-)}
\end{array}\right)\right|^{2}+2\text{Re}\left[e^{-2i\theta_{\text{LO}}}\left\langle \left(\begin{array}{c}
\vec{u}^{(+)}\\
\vec{v}^{(+)}
\end{array}\right)^{*},\left(\begin{array}{c}
\vec{v}^{(-)}\\
\vec{u}^{(-)}
\end{array}\right)\right\rangle \right]\label{eq:Y-variance_Appendix}
\end{eqnarray}
\end{widetext}
where $\langle\cdot,\cdot\rangle$ denotes the inner product between
vectors in $\mathbb{C}^{n}$ and we have defined the vectors
\begin{eqnarray}
{[}\vec{u}^{(\pm)}{]}_{i}&\equiv& U_{i}(\pm\Omega)\sqrt{n_{i}(\pm\Omega+\omega_{\text{d},i})+1/2},\\
{[}\vec{v}^{(\pm)}{]}_{i}&\equiv& V_{i}(\pm\Omega)\sqrt{n_{i}(\mp\Omega+\omega_{\text{d},i})+1/2}.
\end{eqnarray}
As a side remark, we note that $P(\Omega)$ as given by Eq.~(\ref{eq:P-def_Appendix})
coincides with the definition given in the main text as can be seen
using the first equality in Eq.~(\ref{eq:Y-variance_Appendix}) and the
commutator $[\hat{\mathcal{N}}_{\theta_{\text{LO}}}(\Omega),\hat{\mathcal{N}}_{\theta_{\text{LO}}}^{\dagger}(\Omega')]=-|t_{\text{s},\theta_{\text{LO}}}(\Omega)|^{2}\delta(\Omega-\Omega')$.
The Cauchy-Schwarz inequality on $\mathbb{C}^{n}$ implies that
\begin{equation}
\left|\left\langle \left(\begin{array}{c}
\vec{u}^{(+)}\\
\vec{v}^{(+)}
\end{array}\right)^{*},\left(\begin{array}{c}
\vec{v}^{(-)}\\
\vec{u}^{(-)}
\end{array}\right)\right\rangle \right|\leq\left|\left(\begin{array}{c}
\vec{u}^{(+)}\\
\vec{v}^{(+)}
\end{array}\right)\right|\cdot\left|\left(\begin{array}{c}
\vec{v}^{(-)}\\
\vec{u}^{(-)}
\end{array}\right)\right|,\label{eq:Cauchy-Schwarz}
\end{equation}
which leads us to an upper bound of Eq.~(\ref{eq:Y-variance_Appendix})
\begin{equation}
\langle\hat{Y}_{\text{n}}(\Omega)\hat{Y}_{\text{n}}(\Omega')\rangle\leq\left(\left|\left(\begin{array}{c}
\vec{u}^{(+)}\\
\vec{v}^{(+)}
\end{array}\right)\right|+\left|\left(\begin{array}{c}
\vec{v}^{(-)}\\
\vec{u}^{(-)}
\end{array}\right)\right|\right)^{2}.\label{eq:Y-var_bound}
\end{equation}
Note that (for $\Omega,\Omega'>0$)
\begin{multline}
\left|\left(\begin{array}{c}
\vec{u}^{(\pm)}\\
\vec{v}^{(\pm)}
\end{array}\right)\right|^{2}\delta(\Omega-\Omega')  \\
=  \frac{\langle\hat{\mathcal{F}}^{\dagger}(\pm\Omega)\hat{\mathcal{F}}(\pm\Omega')\rangle+\langle\hat{\mathcal{F}}(\pm\Omega)\hat{\mathcal{F}}^{\dagger}(\pm\Omega')\rangle}{2} \\
=  \left[\eta(\pm\Omega)N(\pm\Omega)+\frac{1\mp\eta(\pm\Omega)}{2}\right]\delta(\Omega-\Omega'),\label{eq:uv-vec_length}
\end{multline}
since from the bosonic commutation relations and Eq.~(\ref{eq:scat-rel}) in the main text
we have (for $\Omega,\Omega'>0$)
\begin{multline}
[\hat{a}_{\text{out,e}}(\pm\Omega),\hat{a}_{\text{out,e}}^{\dagger}(\pm\Omega')]=\delta(\Omega-\Omega')\\
\Rightarrow[\hat{\mathcal{F}}(\pm\Omega)\hat{\mathcal{F}}^{\dagger}(\pm\Omega')]=[1\mp\eta(\pm\Omega)]\delta(\Omega-\Omega').
\end{multline}
Combining Eq.~(\ref{eq:P-def_Appendix}) with Eqs.~(\ref{eq:Y-var_bound})
and (\ref{eq:uv-vec_length}) we arrive at the upper bound for $P_{\text{s}}$
given as Eq.~(\ref{eq:P-peak_bound}) in the main text.

\bibliography{alpha}
\bibliographystyle{apsrev4-1}

\end{document}

%% file: generic-transducer_v9.pdf_tex
\begingroup%
  \makeatletter%
  \providecommand\color[2][]{%
    \errmessage{(Inkscape) Color is used for the text in Inkscape, but the package 'color.sty' is not loaded}%
    \renewcommand\color[2][]{}%
  }%
  \providecommand\transparent[1]{%
    \errmessage{(Inkscape) Transparency is used (non-zero) for the text in Inkscape, but the package 'transparent.sty' is not loaded}%
    \renewcommand\transparent[1]{}%
  }%
  \providecommand\rotatebox[2]{#2}%
  \ifx\svgwidth\undefined%
    \setlength{\unitlength}{366.77311454bp}%
    \ifx\svgscale\undefined%
      \relax%
    \else%
      \setlength{\unitlength}{\unitlength * \real{\svgscale}}%
    \fi%
  \else%
    \setlength{\unitlength}{\svgwidth}%
  \fi%
  \global\let\svgwidth\undefined%
  \global\let\svgscale\undefined%
  \makeatother%
  \begin{picture}(1,0.42073005)%
    \put(0,0){\includegraphics[width=\unitlength,page=1]{generic-transducer_v9.pdf}}%
    \put(0.6411313,0.05928106){\color[rgb]{0,0,0}\makebox(0,0)[lt]{\begin{minipage}{0.15482751\unitlength}\raggedright $\omega_{\text{d},a}$\end{minipage}}}%
    \put(0.80662873,0.05928181){\color[rgb]{0,0,0}\makebox(0,0)[lt]{\begin{minipage}{0.5191221\unitlength}\raggedright $\omega_{\text{d},b}$\end{minipage}}}%
    \put(0.52062081,0.26867484){\color[rgb]{0,0,0}\makebox(0,0)[lt]{\begin{minipage}{0.38388861\unitlength}\raggedright $\hat{a}_{\text{in},s}$\end{minipage}}}%
    \put(0.52062081,0.14652847){\color[rgb]{0,0,0}\makebox(0,0)[lt]{\begin{minipage}{0.41660639\unitlength}\raggedright $\hat{a}_{\text{out},s}$\end{minipage}}}%
    \put(0.88477325,0.14652897){\color[rgb]{0,0,0}\makebox(0,0)[lt]{\begin{minipage}{0.38388861\unitlength}\raggedright $\hat{a}_{\text{in},e}$\end{minipage}}}%
    \put(0.88516272,0.26867465){\color[rgb]{0,0,0}\makebox(0,0)[lt]{\begin{minipage}{0.41660639\unitlength}\raggedright $\hat{a}_{\text{out},e}$\end{minipage}}}%
    \put(0.78699578,0.3466614){\color[rgb]{0,0,0}\makebox(0,0)[lt]{\begin{minipage}{0.3555332\unitlength}\raggedright $\hat{\mathcal{F}}$\end{minipage}}}%
    \put(0.4701809,0.41126977){\color[rgb]{0,0,0}\makebox(0,0)[lt]{\begin{minipage}{0.05234845\unitlength}\raggedright b)\end{minipage}}}%
    \put(0.51820109,0.35377507){\color[rgb]{0,0,0}\makebox(0,0)[lt]{\begin{minipage}{0.17578384\unitlength}\raggedright \footnotesize Transducer\end{minipage}}}%
    \put(0,0){\includegraphics[width=\unitlength,page=2]{generic-transducer_v9.pdf}}%
    \put(0.00348026,0.41126985){\color[rgb]{0,0,0}\makebox(0,0)[lt]{\begin{minipage}{0.05234845\unitlength}\raggedright a)\end{minipage}}}%
    \put(0.01575901,0.326336){\color[rgb]{0,0,0}\makebox(0,0)[lt]{\begin{minipage}{0.08579328\unitlength}\raggedright $\omega$\end{minipage}}}%
    \put(0.09600523,0.24474568){\color[rgb]{0,0,0}\makebox(0,0)[lt]{\begin{minipage}{0.35383854\unitlength}\raggedright $l_{\text{a}}\omega_{\text{d,a}}$\end{minipage}}}%
    \put(0.3406471,0.24193914){\color[rgb]{0,0,0}\makebox(0,0)[lt]{\begin{minipage}{0.3629596\unitlength}\raggedright $l_{\text{b}}\omega_{\text{d,b}}$\end{minipage}}}%
    \put(0,0){\includegraphics[width=\unitlength,page=3]{generic-transducer_v9.pdf}}%
    \put(0.09702777,0.09355255){\color[rgb]{0,0,0}\makebox(0,0)[lt]{\begin{minipage}{0.36539187\unitlength}\raggedright $l_{\text{c}}\omega_{\text{d,c}}$\end{minipage}}}%
    \put(0,0){\includegraphics[width=\unitlength,page=4]{generic-transducer_v9.pdf}}%
  \end{picture}%
\endgroup%

%% file: Example-EM-transducer_v1.pdf_tex
\begingroup%
  \makeatletter%
  \providecommand\color[2][]{%
    \errmessage{(Inkscape) Color is used for the text in Inkscape, but the package 'color.sty' is not loaded}%
    \renewcommand\color[2][]{}%
  }%
  \providecommand\transparent[1]{%
    \errmessage{(Inkscape) Transparency is used (non-zero) for the text in Inkscape, but the package 'transparent.sty' is not loaded}%
    \renewcommand\transparent[1]{}%
  }%
  \providecommand\rotatebox[2]{#2}%
  \ifx\svgwidth\undefined%
    \setlength{\unitlength}{210.75bp}%
    \ifx\svgscale\undefined%
      \relax%
    \else%
      \setlength{\unitlength}{\unitlength * \real{\svgscale}}%
    \fi%
  \else%
    \setlength{\unitlength}{\svgwidth}%
  \fi%
  \global\let\svgwidth\undefined%
  \global\let\svgscale\undefined%
  \makeatother%
  \begin{picture}(1,0.44210392)%
    \put(0,0){\includegraphics[width=\unitlength,page=1]{Example-EM-transducer_v1.pdf}}%
    \put(0.62763862,0.0829339){\makebox(0,0)[lb]{\smash{$C(\delta \hat{x})$}}}%
    \put(0,0){\includegraphics[width=\unitlength,page=2]{Example-EM-transducer_v1.pdf}}%
    \put(0.09228746,0.22247221){\color[rgb]{0,0,0}\makebox(0,0)[lb]{\smash{tx line}}}%
    \put(0.12036518,0.1648932){\makebox(0,0)[lb]{\smash{$\gamma_{\text{tx}}$}}}%
    \put(0.47109811,0.28855015){\color[rgb]{0,0,0}\makebox(0,0)[lb]{\smash{+}}}%
    \put(0.48001293,0.09989293){\color[rgb]{0,0,0}\makebox(0,0)[lb]{\smash{--}}}%
    \put(0.16278097,0.30379667){\makebox(0,0)[lb]{\smash{$L$}}}%
    \put(0,0){\includegraphics[width=\unitlength,page=3]{Example-EM-transducer_v1.pdf}}%
    \put(0.69516523,0.1514836){\makebox(0,0)[lb]{\smash{$g$}}}%
    \put(0,0){\includegraphics[width=\unitlength,page=4]{Example-EM-transducer_v1.pdf}}%
    \put(0.6994126,0.35630835){\makebox(0,0)[lb]{\smash{$\gamma_{\text{m}}$}}}%
    \put(1.01645907,0.41370152){\color[rgb]{0,0,0}\makebox(0,0)[lt]{\begin{minipage}{0.04893238\unitlength}\raggedright \end{minipage}}}%
    \put(0.86800691,0.23775734){\makebox(0,0)[lb]{\smash{$\gamma_{\text{wg}}$}}}%
    \put(0,0){\includegraphics[width=\unitlength,page=5]{Example-EM-transducer_v1.pdf}}%
    \put(0.51642866,0.29756913){\makebox(0,0)[lb]{\smash{$\delta \hat{Q}$}}}%
    \put(0.42961723,0.1908269){\makebox(0,0)[lb]{\smash{$\delta \hat{x}$}}}%
    \put(0.01360361,0.02832736){\makebox(0,0)[lb]{\smash{$\hat{b}_{\text{in}},\hat{b}_{\text{out}}$}}}%
    \put(0.87258932,0.11395904){\makebox(0,0)[lb]{\smash{$\hat{a}_{\text{in}},\hat{a}_{\text{out}}$}}}%
    \put(0,0){\includegraphics[width=\unitlength,page=6]{Example-EM-transducer_v1.pdf}}%
    \put(0.80004448,0.35053427){\color[rgb]{0,0,0}\makebox(0,0)[lt]{\begin{minipage}{0.20840063\unitlength}\raggedright phononic waveguide\end{minipage}}}%
    \put(0,0){\includegraphics[width=\unitlength,page=7]{Example-EM-transducer_v1.pdf}}%
  \end{picture}%
\endgroup%

%% file: homo-heterodyne_v4.pdf_tex
\begingroup%
  \makeatletter%
  \providecommand\color[2][]{%
    \errmessage{(Inkscape) Color is used for the text in Inkscape, but the package 'color.sty' is not loaded}%
    \renewcommand\color[2][]{}%
  }%
  \providecommand\transparent[1]{%
    \errmessage{(Inkscape) Transparency is used (non-zero) for the text in Inkscape, but the package 'transparent.sty' is not loaded}%
    \renewcommand\transparent[1]{}%
  }%
  \providecommand\rotatebox[2]{#2}%
  \ifx\svgwidth\undefined%
    \setlength{\unitlength}{401.97425559bp}%
    \ifx\svgscale\undefined%
      \relax%
    \else%
      \setlength{\unitlength}{\unitlength * \real{\svgscale}}%
    \fi%
  \else%
    \setlength{\unitlength}{\svgwidth}%
  \fi%
  \global\let\svgwidth\undefined%
  \global\let\svgscale\undefined%
  \makeatother%
  \begin{picture}(1,0.43750557)%
    \put(0,0){\includegraphics[width=\unitlength,page=1]{homo-heterodyne_v4.pdf}}%
    \put(0.12540278,0.37045646){\color[rgb]{0,0,0}\makebox(0,0)[lt]{\begin{minipage}{0.10733026\unitlength}\raggedright T\end{minipage}}}%
    \put(0,0){\includegraphics[width=\unitlength,page=2]{homo-heterodyne_v4.pdf}}%
    \put(0.31169669,0.32349344){\color[rgb]{0,0.50196078,0}\makebox(0,0)[lt]{\begin{minipage}{0.17165278\unitlength}\raggedright LO\end{minipage}}}%
    \put(0,0){\includegraphics[width=\unitlength,page=3]{homo-heterodyne_v4.pdf}}%
    \put(0.20348081,0.4354409){\color[rgb]{0,0,0}\makebox(0,0)[lt]{\begin{minipage}{0.40487667\unitlength}\raggedright Asymmetric BS\end{minipage}}}%
    \put(0,0){\includegraphics[width=\unitlength,page=4]{homo-heterodyne_v4.pdf}}%
    \put(0.30796511,0.29364078){\color[rgb]{0,0.50196078,0}\makebox(0,0)[lt]{\begin{minipage}{0.59705316\unitlength}\raggedright $\alpha=|\alpha|e^{i\theta_{\text{LO}}}$\end{minipage}}}%
    \put(0,0){\includegraphics[width=\unitlength,page=5]{homo-heterodyne_v4.pdf}}%
    \put(0.42087272,0.09982234){\color[rgb]{0,0,0}\makebox(0,0)[lt]{\begin{minipage}{0.22037417\unitlength}\raggedright freq.\end{minipage}}}%
    \put(0.18047221,0.04245765){\color[rgb]{0,0.50196078,0}\makebox(0,0)[lt]{\begin{minipage}{0.58997269\unitlength}\raggedright $\omega_{\text{LO}}=\omega_{0,\text{e}}$\end{minipage}}}%
    \put(0.00447421,0.441736){\color[rgb]{0,0,0}\makebox(0,0)[lt]{\begin{minipage}{0.10733026\unitlength}\raggedright a)\end{minipage}}}%
    \put(0,0){\includegraphics[width=\unitlength,page=6]{homo-heterodyne_v4.pdf}}%
    \put(0.24960135,0.16136231){\color[rgb]{0,0,0}\makebox(0,0)[lt]{\begin{minipage}{0.09822603\unitlength}\raggedright $\Omega$\end{minipage}}}%
    \put(0.19627063,0.39777722){\color[rgb]{0,0,0}\makebox(0,0)[lt]{\begin{minipage}{0.59705316\unitlength}\raggedright $\hat{a}_{\text{out,e}}$\end{minipage}}}%
    \put(0.00222835,0.39777722){\color[rgb]{0,0,0}\makebox(0,0)[lt]{\begin{minipage}{0.59705316\unitlength}\raggedright $\hat{a}_{\text{in,s}}$\end{minipage}}}%
    \put(0,0){\includegraphics[width=\unitlength,page=7]{homo-heterodyne_v4.pdf}}%
    \put(0.58824548,0.43541129){\color[rgb]{0,0,0}\makebox(0,0)[lt]{\begin{minipage}{0.10733026\unitlength}\raggedright b)\end{minipage}}}%
    \put(0,0){\includegraphics[width=\unitlength,page=8]{homo-heterodyne_v4.pdf}}%
    \put(0.7831732,0.32030146){\color[rgb]{0,0,0}\makebox(0,0)[lt]{\begin{minipage}{0.10733026\unitlength}\raggedright T\end{minipage}}}%
    \put(0,0){\includegraphics[width=\unitlength,page=9]{homo-heterodyne_v4.pdf}}%
    \put(0.7831732,0.14118551){\color[rgb]{0,0,0}\makebox(0,0)[lt]{\begin{minipage}{0.10733026\unitlength}\raggedright T\end{minipage}}}%
    \put(0,0){\includegraphics[width=\unitlength,page=10]{homo-heterodyne_v4.pdf}}%
    \put(0.16377496,0.16136231){\color[rgb]{0,0,0}\makebox(0,0)[lt]{\begin{minipage}{0.09822603\unitlength}\raggedright $-\Omega$\end{minipage}}}%
  \end{picture}%
\endgroup%